\documentclass[10pt,letterpaper]{article}
\usepackage[top=0.85in,left=0.75in,footskip=0.75in,marginparwidth=2in]{geometry}

\usepackage[utf8]{inputenc}

\usepackage{cite}

\usepackage{microtype}
\DisableLigatures[f]{encoding = *, family = * }

\usepackage{changepage}

\usepackage[aboveskip=1pt,labelfont=bf,labelsep=period,singlelinecheck=off]{caption}

\makeatletter
\renewcommand{\@biblabel}[1]{\quad#1.}
\makeatother

\usepackage{lastpage,fancyhdr,graphicx}
\usepackage{epstopdf}
\fancyheadoffset[L]{2.25in}
\fancyfootoffset[L]{2.25in}

\usepackage{color}

\definecolor{Gray}{gray}{.25}

\usepackage{graphicx}

\usepackage{sidecap}

\usepackage{wrapfig}
\usepackage[pscoord]{eso-pic}
\usepackage[fulladjust]{marginnote}
\reversemarginpar

\begin{document}
\vspace*{0.35in}

% title goes here:
\begin{flushleft}
{\Large
\textbf\newline{Circular Dichroism Enhancement in Plasmonic Nanorod Metamaterials}
}
\newline
% authors go here:
\\
D.~Vestler,\textsuperscript{1} 
I.~Shishkin,\textsuperscript{2,*}
E.~A.~Gurvitz,\textsuperscript{3}
M.~E.~Nasir,\textsuperscript{4}
A.~Ben-Moshe,\textsuperscript{1}
A.~P.~Slobozhanyuk,\textsuperscript{3,5}
A.~V.~Krasavin,\textsuperscript{4}
T.~Levi-Belenkova,\textsuperscript{1}
A.~S.~Shalin,\textsuperscript{3}
P.~Ginzburg,\textsuperscript{2,3}
G.~Markovich,\textsuperscript{1}
 and A.~V.~Zayats \textsuperscript{4}

\bigskip
\bf{1} School of Chemistry, Raymond and Beverly Sackler Faculty of Exact Sciences, Tel Aviv University, Tel Aviv 6997801, Israel 
\\
\bf{2} School of Electrical Engineering, Tel Aviv University, Tel Aviv, 6997801, Israel 
\\
\bf{3} ITMO University, St. Petersburg 197101, Russia 
\\
\bf{4} Department of Physics, King's College London, Strand, London WC2R 2LS, United Kingdom 
\\
\bf{5} Nonlinear Physics Center, Research School of Physics and Engineering, Australian National University, Canberra, ACT 0200, Australia
\\
\bigskip
* ivanshishkin@post.tau.ac.il

\end{flushleft}

\section*{Abstract}
Optical activity is a fundamental phenomenon originating from the chiral nature of crystals and molecules. While intrinsic chiroptical responses of ordinary chiral materials to circularly polarized light are relatively weak, they can be enhanced by specially tailored nanostructures. Here, nanorod metamaterials, comprising a dense array of vertically aligned gold nanorods, is shown to provide significant enhancement of the circular dichroism response of an embedded material. A nanorod composite, acting as an artificial uniaxial crystal, is filled with chiral mercury sulfide nanocrystals embedded in a transparent polymer. The nanorod based metamaterial, being inherently achiral, enables optical activity enhancement or suppression. Unique properties of inherently achiral structures to tailor optical activities pave a way for flexible characterization of optical activity of molecules and nanocrystal-based compounds.

% now start line numbers

% the * after section prevents numbering
\section*{Introduction}
Optical activity in bulk materials manifests itself via different responses to right and left circularly polarized light. Circular dichroism (CD) \cite{1} is a particular yet very important case, where a material has different absorption coefficients for light of opposite handedness. This bulk characteristic reflects inherent topological properties of a matter on microscopic scales. In particular, chirality in molecules originates from absence of mirror symmetry of electronic wavefunctions and corresponding selection rules, resulting in different interactions with left and right circularly polarized light. This inherent physical behavior has major implications in biological functionalities and, as a result, it is frequently used as a sensing parameter for identification and separation of enantiomers in racemic mixtures. Unfortunately, circular dichroism has relatively small optical signature in comparison with dominant linear polarization effects, 3--4 orders of magnitude smaller than unpolarized absorption in biomolecules \cite{1}. This makes the detection of circular dichroism response from low quantities of analytes even more challenging. 

Some of the major efforts in overcoming these challenges are focused on engineering the enhancement of circular dichroism response with nanostructures which enable tailoring and enhancing optical near-fields. From this point of view, plasmonic-based nanostructures, enable the field enhancement and confinement on the nanoscale \cite{2,3} and already led to variety of sensitive spectroscopic techniques, such as surface-enhanced Raman scattering \cite{5,6,7} and fluorescence \cite{8,9}. 
Tailoring a CD response with nanostructured materials follows the same approach. A large number of metamaterials \cite{11,12} and metametial-inspired plasmonic structures \cite{13} demonstrated differential absorption for left- and right-circularly polarized light. Chiral plasmonic structures have been employed for ultrasensitive detection and characterization of biomolecules \cite{14,15}. CD enhancement has also been demonstrated experimentally with the help of superchiral light \cite{16} and plasmonic nanoparticles \cite{17,18}. The possibility of CD enhancement by nonchiral structures has been theoretically discussed for intrincically nonchiral \cite{19,20,21,Schaferling_OpEx} and chiral plasmonic nanoparticles \cite{Schaferling_PRX} and nanoantennas \cite{22}, high-permittivity silicon nanoparticles supporting magnetic resonances \cite{23,Dionne_CD} and dispersive media with slow-light \cite{24}. 

The use of achiral tailored nanostructures has advantages over chiral nanostructures. For example, the 'superchiral light' approach requires tight alignment of optical setups and the enhancement of optical activity is achieved within spatially small volumes of space \cite{25}. Sensing enhancement with chiral nanoplasmonic antennas requires very accurate calibration procedures, since inherent optical activity in the nanoantennas themselves dominates the response of tested chiral molecule solutions. Consequently, large-scale achiral substrates, capable of optical activity enhancement, are highly beneficial for a wide range of applications. The basic principle behind this type of enhancement could be qualitatively understood from considering semi-classical description of chiral interactions. Optical activity in molecules originates from simultaneous presence of electric and magnetic dipolar response, whose vectorial components should be as collinear as possible (corresponding matrix element is proportional to the cosine of the angle between electric and magnetic dipoles). Consequently, either electric or magnetic local field enhancements (or even better both simultaneously) could substantially enhance the efficiency of optical rotation.

Various plasmonic structures (antennas) were shown to deliver high electric Purcell enhancements via small modal volumes, e.g. \cite{26,27}. (Magnetic Purcell factors could be achieved with high-index dielectric particles \cite{23,28}). However, in the majority of isolated plasmonic antenna-based configurations local field enhancement is achieved in small spatial regions. At the same time, molecules, statistically spread within the volume, have reduced probability to experience such enhancement, and, as the result, overall improvement of a signal is relatively low. On the other hand, large-scale metamaterials, utilizing coupling between adjacent unit cells, could provide high field enhancement over large areas \cite{wurtz -opt-exp}. Such gold nanorod arrays exhibit transverse and longitudinal plasmonic resonances.  In case of small aspect ratio of the rods (length of the rod is no larger than 10 diameters), the transverse and longitudinal modes hybridize. The strong coupling between the dipolar resonances supported by gold nanorods (corresponding to longitudinal modes) leads to the formation of extended (guided) hybrid mode of the array of similar polarization.
In particular, arrays of free standing vertically aligned nanorods were recently shown to exhibit strong optical anisotropy \cite{36}, deliver record high Purcell enhancement of luminescence signals, generated by fluorescent dye solutions, introduced within the space between the rods \cite{29}. The basic mechanism behind this type of enhancement is an interplay between local near-field interactions and macroscopically averaged enhancement within the composite \cite{30}.

Here we demonstrate that the gold nanorod metamaterial can be used as an efficient platform for circular dichroism enhancement. By tailoring the collective response of the nanorods, it is possible to improve the CD signal of optically active nanocrystals (NCs) by a factor of two.

\section*{Experimental methods}
\textit{Metamaterial fabrication.} Plasmonic nanorod metamaterials (Figs. \ref{fig1}, \ref{fig2}) were fabricated by Au electrodeposition into highly ordered nanoporous anodic alumina oxide (AAO) templates on glass cover slips \cite{35,36}. An Al film of 700 nm thickness was deposited on a substrate by magnetron sputtering. The substrate comprised a glass cover slip with a 10 nm thick adhesive layer of tantalum pentoxide and a 7 nm thick Au film, acting as a weakly conducting layer. Highly ordered, nanoporous AAO was synthesized by a two step-anodization in 0.3M oxalic acid at 40 V. After an initial anodization process, the formed porous layer was partly removed by etching in a solution of H$_{3}$PO$_{4}$ (3.5$\%$) and CrO$_{3}$ (20 g L$^{-1}$) at 70 $^{\circ}$C, which resulted in an ordered pattern. At the next step, the sample was anodized again under the same conditions as in the first step. The anodized AAO was subsequently etched in 30 mM NaOH to achieve pore widening. Gold electrodeposition was performed with a three electrode system, using a non-cyanide solution. Free standing gold nanorod metamaterials were obtained after dissolving the nanoporous alumina template in a mix solution of 0.3 M NaOH and 99.50$\%$ ethanol. The length of nanorods was controlled by the electrodeposition time. Small aspect ratio rods were used in order to reduce absorption in order to avoid artifacts in the CD measurements. Under normal incidence illumination used throughout this work, only one absorption resonance is observed related to plasmonic modes excited with the electric field perpendicular to the nanorod axes [Fig.~\ref{fig1}(b)]. The calculated effective permittivity components of the studied studied meatamaterial are presented in [Fig.~\ref{fig1}(c)]. 

\textit{HgS nanocrystals synthesis and mixing.} Mercury (II) nitrate monohydrate 98.5$\%$, Thioacetamide 98$\%$ and Polyvinyl alcohol 99$\%$ hydrolyzed (146000-186000 Mw) (PVA) were purchased from Sigma-Aldrich, D-penicillamine 99$\%$ was purchased from Fluka, Sodium hydroxide pellets were purchased from Frutarom and Isopropyl Alcohol AR (IPA) was purchased from Gadot. All materials were used without additional treatment. Deionized water were prepared by an USF Elga UHQ water purifier. Nanocrystal synthesis was performed in disposable receptacles rinsed with deionized water.The process is described in detail in \cite{31}. An aqueous solution of Hg(NO$_{3}$)$_{2}$ $\cdot$ H$_{2}$O (100 mM, 0.9 mL) was added to 3 mL of deionized water followed by an aqueous solution of D-penicillamine (0.9 mL, 100 mM), an aqueous solution of NaOH (0.3 mL, 1M) and finally an aqueous solution of Thioacetamide (0.9 mL, 100 mM). The solution was vigorously stirred during the addition and allowed to stir for several hours in the dark at room temperature. The product was separated and washed by adding roughly the same volume of IPA to the solution (the solution turns turbid when shaken once enough has been added) followed by centrifugation (5000 RPM, 5 Minutes) and redispersion in deionized water (this process was repeated 2-3 times).

\textit{Sample fabrication} The HgS NCs solution was concentrated via the centrifugation process described above and mixed in a 1:1 volume ratio with 4$\%$ w/w PVA aqueous solution so the resulting solution would be 10 times more concentrated than the synthesis product. This NCs-polymer mixture was spin-coated on the gold nanorod samples by depositing 50 $\mu$L with the spin-coater (NiLo scientific) set at 3000 RPM for 120 seconds. The typical size of the HgS NCs is in the range of 10-20 nm.

\textit{Circular dichroism measurements.} CD measurements were performed with an Applied Photophysics Chirascan CD spectrometer at normal incidence. The substrates were mounted on a sample holder which can rotate around the normal to the substrate and make sure that the CD spectra do not change with rotation, in order to eliminate linear polarization anisotropy artifacts.  

\textit{Numerical Modeling.} The finite elements method was used for numerical modeling of the effect. The model of the metamaterial with a nanorod of a radius of 30 nm was considered in non-absorbing dielectric medium with refractive index of 1.5 and thickness of 300 nm. Material parameters of gold nanorods were obtained by linear interpolation of experimentally available data \cite{37}. The dielectric slab with the gold rod was placed on a silica substrate in contact with a perfectly matched layer (PML) in order to emulate an infinitely thick substrate. Periodic boundary conditions were applied to this unit cell, corresponding to the inter-rod distance of 100 nm, in order to replicate an infinite array of nanorods. The chiral medium which fills the space between the rods, was introduced via the following constitutive relations: 
\begin{eqnarray}
	\vec{D} = \varepsilon \vec{E} - ik \sqrt{\varepsilon \mu}\vec{H} \label{eq1} \\
	\vec{B} = \mu \vec{H} + ik \sqrt{\varepsilon \mu}\vec{E} \nonumber
\end{eqnarray}
The averaged background permittivity was $\varepsilon$=2.25 and $\mu$=1. Both electric and magnetic susceptibilities were assumed to be isotropic. Dispersion of the Pasteur medium $(k)$ was extracted from the experimental data reported in \cite{31}. 

\section*{Results and discussion}
\begin{figure}[ht]
	\centering
	\includegraphics[width=13cm]{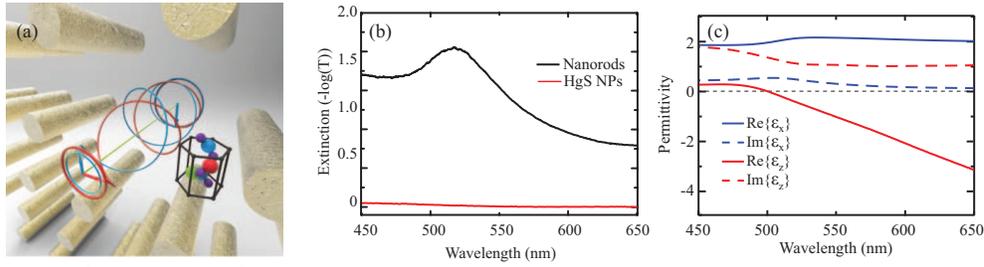}
	\caption{\label{fig1}(a) Schematics of the nanorod metamaterial with a chiral nanocrystal (a hexagonal box) introduced between the nanorods. 
		(b) Normal-incidence extinction spectra measured for the nanorods metamaterial (black line) and HgS nanoparticles (red line). The metamaterial parameters are 60$\pm$10 nm nanorod diameter, 270 nm length, and 100$\pm$20 nm nanorod array period; the nanorods are in air.  
		(c) Effective permittivity components of the nanorod metamaterial calculated using the metamaterial parameters as in (b).}
\end{figure}

\begin{figure}[ht]
	\centering
	\includegraphics[width=13cm]{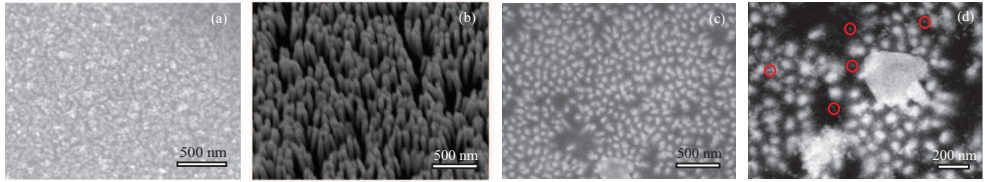}
	\caption{\label{fig2} SEM images of
		(a) HgS nanocrystals drop-casted on silicon substrate,
		(b) nanorod metamaterial before spin-coating of the PVA layer with HgS NCs, 
		(c) metamaterial after the PVA coating with HgS NCs,
		(d) Zoomed view of (c). Red circles mark the positions of individual HgS NCs.}
\end{figure}

We investigated the chiral response of enantio-pure HgS nanocrystals dispersed in a polymer film and introduced between the nanorods. It should be noted that the HgS nanocrystals have an intrinsically chiral crystal structure (space group P$3_{1}21$ or P$3_{2}21$). When they are prepared in the presence of chiral L-penicillamine molecules it was shown that primarily one enantiomorph is formed, which exhibits very large CD signal at the absorption threshold of the nanocrystals at $\sim$550 nm \cite{31}. This enables overlapping the CD resonance of NCs with the resonance of the metamaterial at normal incidence of light [Fig.~\ref{fig1}(b)]. In order to demonstrate the enhancement of circular dichroism, the nanorod metamaterial was infiltrated with a 2$\%$ polyvinyl alcohol (PVA) aqueous solution containing enantio-pure HgS NCs [Fig.~\ref{fig2}]. The contribution of the NCs to overall extinction of the composite is negligible compared to the extinction of the metamaterial alone [Fig.~\ref{fig1}(b)].

\begin{figure}[ht]
	\centering
	\includegraphics[width=13cm]{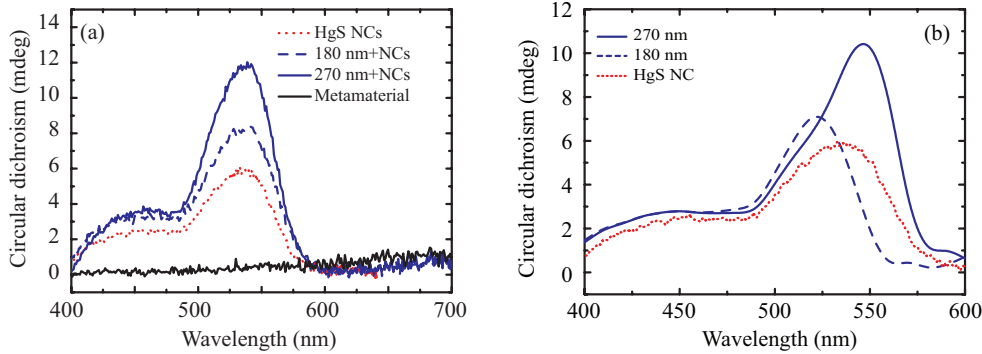}
	\caption{\label{fig3} (a) Experimental circular dichroism spectra of the HgS NCs in a PVA film on a glass substrate (red lines), the metamaterial before (black lines) and after (blue lines) coating with the PVA-HgS NCs film for the nanorod lengths of 180 nm (dashed blue line) and  270 nm (solid blue line). (b) Numerically calculated circular dichroism of the HgS NCs (red lines) and the metamaterial-HgS NCs composite for the nanorod length of 180 nm (dashed blue line) and  270 nm (solid blue line). 
		All other parameters of the metamaterials are as in Fig. 1 (b). Both experiments and simulations are performed at normal incidence. }
\end{figure}

The CD spectrum of the bare nanorods (before coating) shows a weak non-zero CD signal [Fig. \ref{fig3}(a)] which can be attributed to minor tilts of individual rods from the normal to the sample surface, resulting in minor anisotropy of absorption of LCP and RCP light by the samples. On the other hand, CD of HgS NCs embedded in the PVA film on a glass substrate (without the metamaterial) has a pronounced resonance around 525 nm and a wide shoulder at shorter wavelengths. The metamaterial clearly influences the CD of the nanocrystals and the enhancement depends on the nanorod length. The maximum observed enhancement is about two-fold in the case of the longest rods (280 nm). It is important to note that the CD signal, defined as a difference between transmission coefficients of right and left circularly polarized light, normalized by their sum, does not depend on the thickness of a film under investigation. 
In order to verify the repeatability of the observed enhancement a set of control experiments was carried out. The PVA layer with HgS NCs was washed away with a mixture of deionized water, IPA and dimethylsulphoxide heated to 70$^{\circ}$C followed by the CD spectroscopy of the washed sample. Afterwards, a new coating of PVA-HgS NCs was deposited on the cleaned nanorod array followed by CD spectroscopy. The sample was then coated with PVA without HgS NCs in order to check, that no optical activity was introduced by PVA coating. The small changes (< 25 \% in the enhancement factor) between repeated coating cycles are believed to be due to slightly different polymer distribution on the sample surface achieved at each cycle and minor differences in the positioning of the sample in the holder of spectrometer.

\begin{figure}[h!]
	\centering
	\includegraphics[width=12cm]{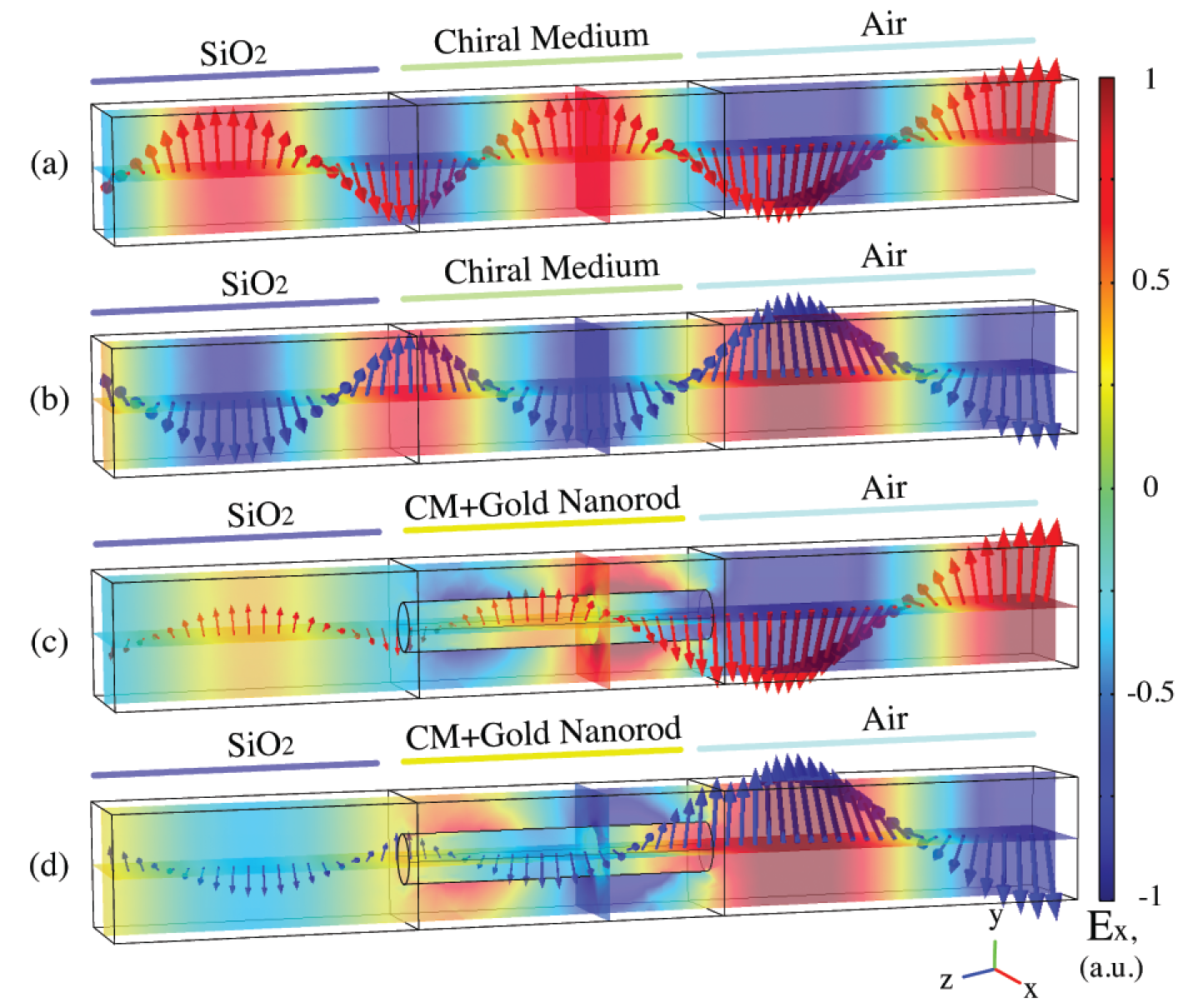}
	\caption{\label{fig4} Simulated electromagnetic wave propagation (from right to left) through (a,b) chiral medium without the metamaterial and (c,d) chiral medium embedded in the metamaterial for (a,c) RCP and (b,d) LCP incident light. The instantaneous field $E_x$ is shown as the color map. Arrows indicate the direction of the electric field with their length being proportional to the logarithm of the instantaneous field amplitude. 
	}
\end{figure}
In order to understand the impact of the metamaterial on the CD enhancement, the propagation of circularly polarized waves in the metamaterial with chiral NCs was numerically studied. The modelled CD response for both the PVA-NCs film on a glass substrate is in good correspondence with the experimental spectra [Fig. \ref{fig3}(b)]. In the case of the metamaterial composite, a very good agreement with the experiment in terms of the enhancement factor and the spectral response is obtained for 270 nm long rods, without any additional fitting parameters. At the same time, for shorter rods (180 nm length), while the same trend is observed as in the experiment, the overall spectral shape is somewhat different (this may attributed to uncertainties in geometrical parameters of the nanorod arrays and the thickness of the polymer layer, spin coated within the free standing rods template).

It is worth noting the alternative approach towards the evaluation of CD enhancement, which relies on calculation of chiral fields of the nanostructures. It is described in detail in \cite{Schaferling_OpEx,Schaferling_PRX,Dionne_CD}, where isolated nanostructures were investigated. Such approach is best suited for evaluation of properties of single isolated nanostructures or their sparse arrays. In order to estimate a collective response from a bi-anisotropic material the chiral field should be integrated over a volume, which it occupies. Remarkably, dipolar resonators do not provide an overall collective enhancement, as was demonstrated in \cite{Schaferling_OpEx}. However the nanorod metamaterial case provides conceptually different scenario, since the collective mode of a structure is formed due to the near-field coupling. Furthermore, being optically thick, it possesses distributed optical losses, which affect chiral contributions from different individual crystals, situated in the vicinity of the rods. As the result, nanorod metamaterial demonstrates peculiar properties of the collective CD enhancement. Detailed comparison between two approaches (chiral field integration and propagation of circularly polarized light in a bulk optically active media) will be covered in a forthcoming report. 

The electric field distribution and its orientation for right (RCP) and left (LCP) circularly polarized waves propagating through the meta-material with and without chiral NCs are shown in Figure \ref{fig4}. The excited plasmonic resonance shown in [Fig.~\ref{fig4}(c,d)] provides different absorption enhancement for RCP and LCP waves propagating through the medium in comparison with the case of chiral medium with no rods introduced [Fig.~\ref{fig4}(a,b)]. It should be noted, that the propagation through metamaterial layer results in additional losses.
\section*{Outlook and Conclusions}
Nanorod metamaterial was demonstrated to be an efficient platform for CD enhancement. When the metamaterial resonance was tuned to spectrally overlap with the chiroptical resonance of the NCs, an overall two-fold CD enhancement was observed. Near-field interaction between the HgS NCs and the nanorods forming the metamaterial dictate the overall CD behavior. Several light-matter interaction processes, tailored by the nanorod metamaterials were previously investigated and local field effects were similarly shown to be an important factor (see, e.g. \cite{38,39}). Collective macroscopic enhancement of microscopic optical activity with large scale self-assembled tailored nanostructures is of a paramount importance for enabling many applications, where accurate optical sensing of circular dichroism is required. 

\section*{Acknowledgments}
This work has been supported, in part, by EPSRC (UK) and the ERC iPLASMM project (321268). P.G. acknowledges the support from TAU Rector Grant, PAZY foundation, and German-Israeli Foundation (GIF, grant 2399). G.M. acknowledges support by The Israel Science Foundation grant no. 507/14. A.B.M. was supported by the Adams Fellowship Program of the Israel Academy of Sciences and Humanities.  E.G. and A.S. acknowledge support of the Russian Fund for Basic Research within the project 16-52-00112. The numerical simulations of electromagnetic field distribution in metasurface have been supported by the Russian Science Foundation Grant No. 16-12-10287. A.S. acknowledges the support of Scholarship SP-4248.2016.1 and the support of Ministry of Education and Science of the Russian Federation (GOSZADANIE Grant No. 3.4982.2017/6.7). A.Z. acknowledges support from the Royal Society and the Wolfson Foundation. The data access statement: All data supporting this research are provided in full in the results section.
\end{document}